\def\zb{$z$-band}
\def\vb{$V$-band}
\def\gtorder{\mathrel{\raise.3ex\hbox{$>$}\mkern-14mu\lower0.6ex\hbox{$\sim$}}}
\def\ltorder{\mathrel{\raise.3ex\hbox{$<$}\mkern-14mu\lower0.6ex\hbox{$\sim$}}}
\def\eg{{\it e.g.~}}
\def\etc{{\it etc.~}}
\def\etal{{\it et al.~}}

\def\asec{^{\prime\prime}}

\documentclass[12pt,preprint]{aastex}
\slugcomment{}

\lefthead{Rix \etal } \righthead{Galaxy Evolution from
Morphologies and SEDs}

\begin{document}

\title{GEMS: Galaxy Evolution from Morphologies and SEDs}

\author{
Hans-Walter Rix\altaffilmark{1}, Marco Barden\altaffilmark{1},
Steven V.W. Beckwith \altaffilmark{2}, Eric F.
Bell\altaffilmark{1}, Andrea Borch\altaffilmark{1}, John A. R.
Caldwell\altaffilmark{2}, Boris H\"au\ss ler\altaffilmark{1}, Knud
Jahnke\altaffilmark{3}, Shardha Jogee\altaffilmark{2}, Daniel H.
McIntosh\altaffilmark{4}, Klaus Meisenheimer\altaffilmark{1},
Chien Y. Peng\altaffilmark{5}, Sebastian F.
Sanchez\altaffilmark{3}, Rachel S. Somerville\altaffilmark{2},
Lutz Wisotzki\altaffilmark{3}, Christian Wolf\altaffilmark{6} }
\affil{1 Max-Planck Institute for Astronomy, D-69117 Heidelberg,
Germany} \affil{2 Space Telescope Science Institute, Baltimore, MD
21218} \affil{3 Astrophysikalisches Institut Potsdam, D-14482
Potsdam, Germany} \affil{4 University of Massachusetts, Amherst,
MA 01003} \affil{5 University of Arizona, Tucson, AZ 85721}
\affil{6 University of Oxford Astrophysics, Oxford OX1 3RH, UK}

% Notice that each of these authors has alternate affiliations, which
% are identified by the \altaffilmark after each name.  The actual alternate
% affiliation information is typeset in footnotes at the bottom of the
% first page, and the text itself is specified in \altaffiltext commands.
% There is a separate \altaffiltext for each alternate affiliation
% indicated above.

% The abstract environment prints out the receipt and acceptance dates
% if they are relevant for the journal style.  For the aasms style, they
% will print out as horizontal rules for the editorial staff to type
% on, so long as the author does not include \received and \accepted
% commands.  This should not be done, since \received and \accepted dates
% are not known to the author.

\begin{abstract}
GEMS, {\bf G}alaxy {\bf E}volution from {\bf M}orphologies and
 {\bf S}EDs, is a large-area (800 arcmin$^2$) two-color (F606W and F850LP)
imaging survey with the Advanced Camera for Surveys on HST.
Centered on the Chandra Deep Field South, it covers an area of
$\sim 28^\prime \times 28^\prime$, or about 120 Hubble Deep Field
areas, to a depth of m$_{AB}$(F606W)$=28.3 (5\sigma )$ and
m$_{AB}$(F850LP)$=27.1 (5\sigma )$ for compact sources.
 In its central $\sim 1/4$, GEMS
incorporates ACS imaging from the GOODS project. Focusing on the
redshift range $0.2\ltorder z \ltorder 1.1$,
GEMS provides morphologies and structural
parameters for nearly $10,000$ galaxies where redshift estimates,
luminosities and SEDs exist from COMBO-17. At the same time, GEMS
contains detectable host galaxy images for several hundred faint AGN.
This paper provides an overview of the science goals, the
experiment design, the data reduction and the science analysis
plan for GEMS.
\end{abstract}

% The different journals have different requirements for keywords.  The
% keywords.apj file, found on aas.org in the pubs/aastex-misc directory,
% contains a list of keywords used with the ApJ and Letters.  These are
% usually assigned by the editor, but authors may include them in their
% manuscripts if they wish.

%\keywords{globular clusters,peanut clusters,bosons,bozos}

% That's it for the front matter.  On to the main body of the paper.
% We'll only put in tutorial remarks at the beginning of each section
% so you can see entire sections together.

% In the first two sections, you should notice the use of the LaTeX \cite
% command to identify citations.  The citations are tied to the
% reference list via symbolic KEYs.  We have chosen the first three
% characters of the first author's name plus the last two numeral of the
% year of publication.  The corresponding reference has a \bibitem
% command in the reference list below.
%
% Please see the AASTeX manual for a more complete discussion on how to make
% \cite-\bibitem work for you.

\section{Introduction}

The formation and the evolution of galaxies are driven by two
interlinked processes. On the one hand, there is the dynamical
assembly of the mass distribution in the context of dark-matter
dominated, hierarchical structure formation. On the other hand,
there is the star-formation history (SFH), i.e. the successive
conversion of gas into stars, along with the ensuing
feedback. By now, the cosmological background model and the
corresponding large-scale structure growth are well constrained
(\eg Percival \etal 2002; Spergel \etal 2003; we will use
$\Omega_{M}=0.3$, $\Omega_\Lambda=0.7$ and $H_0=70$ km/s
throughout), and the focus is shifting towards understanding on
galaxy scales the dynamics and the physics of star formation,
reflected in the structure and stellar populations of 
the resulting galaxies.

On the one hand, a comprehensive picture of galaxy formation 
must match the population statistics of
integrated galaxy properties, \eg the galaxy luminosity
and mass functions
or the overall spectral energy distributions (SED),
and the dependence of these distributions on the larger environment.
But a picture of galaxy
formation should also explain the internal structure of galaxies: e.g.
their size, bulge-to-disk ratio, degree of symmetry, internal
population gradients, and nuclear properties. Many of the
ingredients that determine the internal structure and the SFH are
qualitatively clear. For example, the size of galaxies is linked to the angular
momentum of the stars and their progenitor gas,  created early on
through tidal torquing; spheroid stars formed before or during the last
episode of violent relaxation, whereas most disk stars in large
galaxies have formed after the last major merger; and major mergers
are effective triggers of powerful starbursts, if the progenitor
galaxies have a sizeable supply of cold gas. These same mergers
are also suspected to trigger nuclear (AGN) activity by funneling
gas into the vicinity of the ubiquitous central black holes.

Quantitative theoretical predictions of the resulting
internal structure and SFHs of individual galaxies are neither 
robust nor unique, as galaxy formation involves a vast range of spatial scales, from 
well below 1~pc to well above 1~Mpc, 
along with complex geometries. Neither direct
numerical simulations (\eg Katz \& Gunn 1991; Steinmetz \&
Navarro 2002; Springel, Yoshida, \& White 2001), nor semi-analytic models
(\eg Cole \etal 2000; Kauffmann, White, \& Guiderdoni 1993; 
Somerville \& Primack 1999)
can currently capture all important aspects of the problem.
Turning to empiricism in light of this situation,
galaxy evolution is perhaps best studied by observing directly how 
the properties of the galaxy population change with cosmic epoch.

Observational constraints on the galaxy population in the
present-day ($z\ltorder 0.2$) ~universe have drastically improved
over the last years, in particular through three large surveys:
2MASS imaging the sky in the near-infrared (Skrutskie \etal 1997), 
and the optical surveys SDSS (York \etal 2000) and 2DFRS
(Colless \etal 2001). The galaxy luminosity functions, the galaxy
(stellar) mass function, the galaxy correlation function, the
distribution of spectral energy distributions, the distribution of
galaxy sizes, \etc have been (re-)determined with unprecedented
accuracy (\eg Blanton \etal 2003c; Norberg \etal 2002; Shen \etal 2003;
Kauffmann \etal 2003; Bell \etal 2003a; Strateva \etal 2001).

The observational challenge now is to come up with a correspondingly 
more detailed
assessment of galaxy properties and galaxy population properties
at earlier epochs. Over the last decade the `look-back' approach
to studying galaxy evolution has experienced a number of
breakthroughs, both in obtaining distant galaxy samples with
secure redshifts to study integrated galaxy
properties, and in obtaining high-resolution imaging, mostly with HST, to
study internal structure
 (\eg Ellis \etal 1996; Lilly \etal 1996; Steidel \etal 1996;
 Williams \etal 1996; van Dokkum \etal 1998).
Yet, for a full comparison with local samples, these data sets -- typically
a few hundred objects -- have been much too small. This holds especially
true considering only samples with redshift {\it and} internal structure information. 
For one, these samples are too small in number to allow dissecting
the galaxy population by redshift, luminosity, color, size or even environment,
and still be left with significant subsamples. Previous samples with redshifts
and well-resolved images have also been drawn from too small an area. 
As a consequence,
they cannot reflect the `cosmic average' at any epoch, because 
luminous galaxies are clustered quite strongly at all epochs
(\eg Giavalisco \etal 1998, Phleps and Meisenheimer, 2003).

Existing studies of morphology and internal structure have shown
that to $z\sim 1$ the sizes and Hubble types of galaxies roughly
resemble the nearby universe (\eg Abraham \etal 1996; Lilly \etal
1998; Simard \etal 1999), whereby the significance of possible
differences from $z\sim 0$ (\eg the higher incidence of distorted
morphologies) is weakened by discrepant sample definitions, small
sample sizes and survey volumes and by the observational effects
of $(1+z)^4$ surface brightness dimming and of bandpass shifting.
At $z\gtorder 2$, galaxy images lose their {\it prima facie}
resemblance to the nearby universe and appear more compact, but
there, too, the band pass shifting may give an exaggerated
impression of true evolutionary effects of the galaxy population
(\eg Labbe \etal 2003).

Recently, the COMBO-17 project (Wolf \etal 2001; Wolf \etal 2003a, 
W03) has afforded a thirty-fold increase by number over the
earlier redshift surveys [See also Fried \etal 2001; Im \etal 2002;
Le Fevre \etal 2003; Davis \etal 2002 for other recent or ongoing 
surveys] . COMBO-17 incorporates deep ($R \la
23.5$) photometry in 17 optical filter bands, providing redshifts
good to $\delta$z$/(1+$z$) \ltorder 0.02$ for both galaxies and AGNs. From a
sample of $\ga 25\,000$ galaxies with $z \la 1.2$, the survey has
explored the population and integrated properties of galaxies
since these redshifts.  Building on earlier results, COMBO-17 has
detailed and quantified the increasing prominence of massive
galaxies without young stars (Bell \etal 2003b), the shift of
high specific star formation activity to low mass systems (W03),
and the SED-differential evolution of galaxy clustering (Phleps \&
Meisenheimer 2003). Furthermore, COMBO-17 has provided a deep insight
into the population and evolution of low-luminosity AGNs (Wolf
\etal 2003b).

Yet, as any other ground-based imaging survey, COMBO-17 could add
little to elucidating the evolution of internal structure over
this redshift interval. The goal of the present project, GEMS 
(Galaxy Evolution from Morphologies and SEDs), is to provide
an order of magnitude improvement
in assessing the evolution of the {\it internal structure and
morphology} of galaxies over the `last half' of cosmic history (actually the
last 8.5~Gyrs to $z\sim 1.2$)
through wide-area, high-resolution imaging with HST.

Foremost, GEMS should address a) how the galaxy
merger and tidal interaction rate evolved since $z\sim1$; b) which
portion of the global star formation rate at any given epoch is
externally triggered, through tidal interaction or mergers; c)
whether stellar disks grew ``inside out"; d) whether the formation
of bulges entirely preceded the formation of their surrounding
disks; e) whether stellar bars are a recent ($z\leq1$) phenomenon;
f) whether the drastic decay in nuclear accretion activity
is reflected in any drastic change of the host galaxy population.

Obviously, we would be most interested in tracing the evolution of
individual objects. Yet, only the evolution of population
properties is observable. In practice, one tries to bridge this gap
and answer the above questions
by assessing separately the redshift evolution of various structural
parameter relations and of space densities for different 
galaxy sub-samples: e.g. the relation of 
disk- or bulge-size {\it vs.} their luminosity or stellar mass; 
the space density of large disks;
the ratio of disk and bulge stars at different epochs, the fraction
of young stars in disks, etc.. 

GEMS, and a number of other cosmological imaging
surveys (in particular the narrower, but deeper GOODS survey;
Giavalisco \etal 2003, Ferguson et al. 2003, Moustakas et al. 2003) with HST, have been
enabled by the advent of the Advanced Camera for Surveys (ACS,
Ford \etal 2003), which vastly improves the survey speed of HST.
The GEMS observations were planned at a time when much research
effort in observational cosmology is centered around a number of
selected fields, such as the HDF's North and South, or the Chandra
Deep Fields, where multiwavelength observations from the X-rays to
the far-IR and radio are creating synergies; the GEMS mosaic
encompasses such a field, the Chandra Deep Field South.

In the remainder of the
paper we outline the experiment design (\S 2), the initial data
reduction (\S 3), the image analysis and initial galaxy catalog (\S 4),
and the planned science analysis.

\section{Experimental  Design}

The immediate goal of the GEMS survey is to provide high resolution
images from which to extract an empirical data base of `structural
parameters' that describe the stellar bodies for a large sample of
distant galaxies. Here we outline the rationale for the particular
survey implementation.

To resolve the internal structure of galaxies at $z\sim 1$, with
expected typical scalelengths of $\sim 2$~kpc, 
one needs imaging 
at a spatial resolution considerably  finer than their apparent 
size: e.g. 2~kpc project to $0.\asec 26$ at $z=0.75$.
To date HST is still far more efficient to deliver this over wide fields over wide fields
than AO on large ground-based telescopes, such as afforded by
CONICA on the VLT (Lenzen \etal 2003).

\noindent{\sl Sample Size:} The {\it desideratum} is the
distributions of galaxy size, light concentration,
bulge-to-disk-ratio, and morphology as a function of redshift,
luminosity, SED, and perhaps environment. Even considering only
one number to characterize the internal structure of galaxies, one
needs to estimate the frequency distribution of galaxies in a
four-dimensional parameter space , ($z$,L,SED,structure). For a handful of bins per axis
and $\sim$10 galaxies per bin (or S/N $\geq 3$), one needs samples
of $\sim 10^4$ galaxies.

\noindent{\sl Choice of Survey Area and  Field: } To approach
representative sampling of environments, the field size must be
well in excess of the correlation length of typical (L$_*$) galaxies,
which is $\sim$ 5~Mpc comoving
for $0.3 \ltorder z \ltorder 1$, Phleps and Meisenheimer 2003; Coil \etal
2003), and even twice as large for red, early type galaxies (e.g. Daddi et al 2001).
This scale corresponds to $7^\prime - 11^\prime$ at $z=0.75$, or
three times HST's field-of-view in a single pointing 
($\sim 3^\prime$) with the ACS (Ford \etal 2003). The need for
large samples with redshifts, faint limiting magnitudes and imaging with HST's
restricted field of view lead to a densely sampled, contiguous
field. To date, the COMBO-17 survey provides redshifts and SEDs in
three disjoint fields (WO3) of $\sim 30' \times 30'$ each, one
including the Chandra Deep Field
South (CDFS, Giacconi \etal 2001), which we refer to as the `extended
CDFS' (E-CDFS) area. Note that the results of W03 show, that even 
for such large field sizes the galaxy population variations due to large scale
structure are still very significant, e.g. $>50$\% {\it rms} for luminous red
galaxies over redshift intervals of $\delta$z$\sim 0.2$.

In part, we chose the E-CDFS because it appeared {\it a priori}
representative with respect to its galaxy population, as opposed
to \eg the Abell 901 cluster field in COMBO-17. But foremost, the field is
preferable because of the intense focus of research at other,
complementary wavelengths, in particular in X-rays with Chandra
and XMM observations (Rosati \etal 2002) and in the thermal
infrared with upcoming SIRTF observations (GOODS, Dickinson \etal
2003). As we will detail in \S3, GEMS imaging is coordinated with
the multi-epoch GOODS imaging over the central $\sim$ 25\% of the
total GEMS area. The GEMS survey area and its spatial relation to
the GOODS and COMBO-17 field is illustrated in
Figure~\ref{gems-layout}. The central co-ordinates of the
COMBO-17, and hence GEMS, field are $\alpha$=03h~32m~25s,
$\delta =-27~48^\prime$~50$^{\prime \prime}$ (2000).

\noindent{\sl Flux Limit and Filters:} To reach `typical' galaxies
(L $\sim L_*$) to redshifts of z$\sim$1, one needs samples with
redshifts to a magnitude limit of $m_R \sim$ 23.5 (WO3). The
GEMS imaging depth was designed to permit robust galaxy model fits for
most objects that are in the COMBO-17 redshift sample, $m_R \leq
23.6$. To get S/N $\geq$ 20 on extended objects near this
magnitude requires about one orbit of exposure time with the ACS
in F850LP.

The structural parameters (de~Jong 1996; Kranz, Slyz, \& Rix 2003) and
morphology (see \eg Rix \& Rieke 1993), especially of late
types galaxies, depend on the observed wavelength. Therefore, one
must study morphology evolution at comparable rest-frame
wavelengths across the explored redshift. The ACS filters chosen
were F606W (between the Johnson V and R bands, hereafter ``V'') and F850LP
(corresponding approximately to, and hereafter referred to as, the z-band).
 For some redshift ranges these
observations provide immediately
galaxy images in the rest-frame B-band ($\sim$ 4500 \AA).
For most redshifts one can reconstruct such a rest-frame
image through pixel-by-pixel interpolation across the two bands, or
through modest extrapolation in the other redshift ranges. The lowest
redshifts of interest, z$\sim$0.2, require a blueward
extrapolation of the observed V-band flux by 10\% in $\lambda$ and
the highest redshifts, z=1.2, a 10\% redward extrapolation of
the z-band flux. For redshifts in between one can interpolate
between the two observed filters; at z$\sim$0.33 our V-band 
corresponds directly to rest-frame B, as does the z-band at
z$\sim$1. This choice of filters also provides consistency with
the GOODS data at the field center.

Observations in two filters are crucial not only for the
reconstruction of the rest-frame B-band, but also for color
information, especially radial gradients, within one galaxy. Given
limited observing time, area and imaging depth were deemed more
important than a third filter. In cycle 11, 125 orbits of HST time (G0-9500, PI: Rix)
were awarded to carry out this program. All these data have no
proprietary period and are freely accessible.

% Authors may indicate to the editorial staff where they would like
% figures and tables to be placed in the manuscript.  This is done with
% either the \placefigure{KEY} or \placetable{KEY} commands.  These
% commands require \label{KEY} commands to be placed appropriately with
% corresponding table and figure captions.  When the manuscript is
% printed a short note is printed on the page where the figure or table
% is to go.  These commands are ignored in the aaspp4 and aas2pp4 styles.

%\placetable{tbl-3}
%\placefigure{fig1}

% In this section, we see the use of the \subsection command to set off
% an independent subsection.  We only have one here; usually there would
% be several.

% We show the use of several of the displayed math environments described
% in the User Guide, and you get a healthy dose of mathematical typesetting
% examples.  Also, observe the use of the LaTeX \label command after the
% \subsection to give a symbolic KEY to the subsection for cross-referencing
% in a \ref command.  LaTeX automatically numbers the sections, equations,
% tables, etc., as it goes, so in general you don't know what number=
% is going to have.  We'll refer to the "hairymath" section a little later.

\section{Data}

The full details of tile lay-out (Figure~\ref{gems-layout}), the
observations, the data reduction, and the data quality assessment
will be given in Caldwell \etal 2004 ({\it in prep.}, C04); here
we provide a brief overview.

\subsection{Observations}

\begin{figure}
\plotone{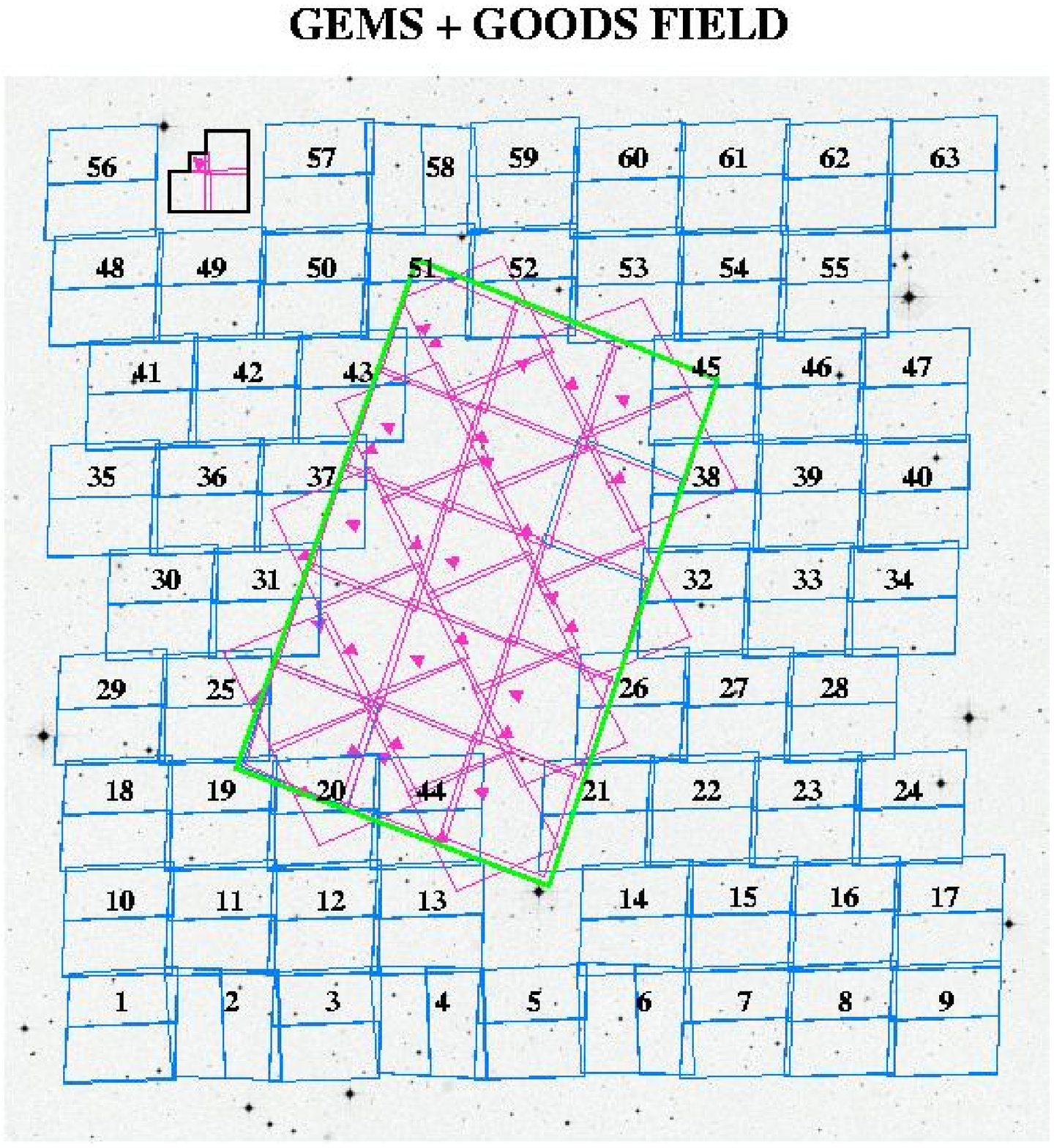} \caption{Layout of the GEMS image
mosaic. With 800 square arcminutes, GEMS 
nearly covers the extended Chandra Deep Field South from
COMBO-17 (underlying r-band image, see W03), which measures $\sim
30^\prime\times 30^\prime$; the orientation is North up and East
left. The individual GEMS tiles, labelled by their HST visit
number are shown as pairs of rectangles (ACS chips). The mosaic
tiles indicated in pink at the center and not aligned with the
overall field are the first epoch observations of GOODS, which
have been incorporated into the overall GEMS analysis. The tilted
large rectangle (solid green line) indicates the area of planned
SIRTF observations for GOODS. A few tiles have been omitted from
the overall mosaic to avoid the brightest stars in the field.
Observations for four tiles (2,4,6,58) had to be at different roll
angles to assure guide stars. The area of the HDFs is indicated at
the top left. \label{gems-layout}}
\end{figure}

The bulk of the GEMS observations (59 visits, or 117 orbits) were
carried out with the ACS's WFC (Ford \etal 2003) between Nov 4
and Nov 24, 2002. Two visits were executed on Sep 14, 2002 and one
each on Feb 24 and Feb 25, 2003. The first epoch observations of
the GOODS survey that cover the central position of the GEMS field
were taken in July and August 2002 (Giavalisco \etal 2003). The
tile pattern of the overall mosaic (Figure~\ref{gems-layout}) was
laid out to a) encompass the GOODS epoch 1 data; b) create a large
contiguous imaging field; and c) avoid excessively bright stars
that would lead to excessive charge bleeding and scattered light
on the CCD. Of the 63 tiles, 59 are oriented North-South. For four
tiles, the availability of guide stars forced an orientation at
right angles to the remaining ones (see Figure 1).

Each HST orbit visit (see labels in Fig.~\ref{gems-layout})
consisted of three separate 12~min to 13~min exposures each for
both V-band (F606W) and z-band (F850LP), dithered by ~3" between exposures. The
exposures of each tile in each filter required one full orbit with
overhead. The dithering was chosen both to close the inter-chip
gap and to provide sub-pixel sampling for drizzling of the final
image. In each visit the first orbit was spent on V and the
second on the z-band, where the rapid re-acquisition allowed a
slightly longer (by 3~min) total exposure time.

\subsection{Data Reduction}

For the first version of the GEMS data the underlying approach was
to reduce each tile in each filter separately, i.e. each set of 3
dithered exposures taken within an orbit was first treated as a
completely independent data set. To assure data homogeneity, we
re-reduced the first epoch GOODS data at the center of the overall
GEMS area in the exactly same way as the GEMS data.

Each frame was processed using {\sl CALACS} (
www.stsci.edu/hst/acs/analysis ) to take care of bias and dark
current subtraction, flatfielding and to include the photometric
calibration information. Frame combination and cosmic-ray
rejection were accomplished with multidrizzle (Mutchler, Koekemoer, 
\& Hack
2003), resulting in a combined image and a variance array on a
0."03/pixel grid (as opposed to the original 0."05/pix of the
individual frames). We opted for a relatively fine 0."03 scale, to
avoid resolution degradation in subsequent operations, even though
it implies more strongly correlated pixel noise (C04).

Cosmic ray rejection with three dithered frames worked
excellently. As GEMS does not address time variable phenomena, any
faint and rare residual cosmic rays are not of concern for its
immediate science goals.

The astrometry of each image tile was tied to the overall catalog
from the ground-based COMBO-17 {\it r}-band  image (Wolf \etal
2001, W03), with an {\it rms\/} of $\sim 0.\asec 14$ per source
(see also \S 4.1). Both filters of each GEMS tile are tied to the
COMBO-17 frame independently, but the V-band frames were
subsequently {\it micro-registered} to a fraction of a pixel
with respect to the z-band frame, for the color distribution analysis
of individual sources.

Flux calibration was done using the best available zero points as
of Feb. 2003, $V_{AB} =26.49$ and $z_{AB} =24.84$. The resulting
point source sensitivities (5 $\sigma$) are: $m_{lim}$ (V) = 28.25
and $m_{lim}$ (z)= 27.10, in AB magnitudes. The angular resolution
of the images, $\lambda$/D$\approx 0.055\asec$ and $0.077\asec$ in
V-band (F606W) and z-band (F850LP), respectively, corresponds to physical
resolutions of 500~pc and 700~pc for galaxies at $z\sim 0.75$,
comparable to galaxies in the Coma Cluster imaged with 1$\asec$ seeing.
Figures~\ref{GEMS-COMBO17} and ~\ref{deep-shallow} give a visual
impression of how the GEMS images compare to the two most
immediately related data sets: the deep {\it r}-band image from
COMBO-17 and the deep, 5-orbit GOODS images. The total
affective area of the GEMS mosaic is 796 arcmin$^2$.

We have not found any significant tile-to-tile variations in the
relevant data properties (noise, sensitivity, \etc ) and it
appears that the intra-tile variations in sensitivity are also
negligible. Further details will be given in C04.

\section{Data Analysis}

As the largest, multi-color image taken with HST to date, GEMS
can be applied to wide range of scientific
problems. Yet, the immediate focus of GEMS is to study the
internal structure of galaxies for which redshifts and SEDs exist
from COMBO-17. To accommodate the narrow and broader goals, the
data analysis is broken down into three steps: 1) a catalog of
``all" objects well detected in the GEMS z-band (F850LP) image, 2) a
match-up with the COMBO-17 catalog, 3) the fitting of
parameterized image models to selected  source postage stamps.
At a later stage this will be followed by the creation
and analysis of color images. As for many applications of
immediate interest, the longest accessible rest-frame wavelength
is most relevant, thus the first version of the GEMS catalog is
``driven" by the  \zb ~image, with the \vb ~image providing color
information.

%\def\plotfiddle#1#2#3#4#5#6#7{\centering \leavevmode
%\vbox to#2{\rule{0pt}{#2}}
%\special{psfile=#1 voffset=#7 hoffset=#6 vscale=#5 hscale=#4 angle=#3}}

%\begin{figure}
%\plotone{hot-cold-SExtraction-v1.eps} \caption{Two-pass strategy for
%object detection and deblending: the left-subpanels show the
%source identification with a conservative (``cold") setting of
%the SExtractor parameters that avoid over-deblending of large
%objects and galaxies with lumpy structure. The right panels shows
%the result of SExtractor with a parameter setting that picks up
%objects closer to the noise threshold, at the expense of
%occasionally breaking up objects erroneously. Our final object
%catalog consistst of the left, cold objects augmented by the
%missing right, ``hot'' objects, but only those that do not overlap
%with a cold object. \label{hot-cold-SExtraction}}
%\end{figure}

\subsection{Object Detection and Deblending}

For parsing an image into an object catalog, the most widely used
image software at present is SExtractor (Bertin \& Arnouts
1996). As for the COMBO-17 catalogs, we apply SExtractor to the
GEMS mosaic to obtain positions and a variety of photometric
parameters for each detected source. We configure SExtractor to
produce a GEMS source catalog that a) contains nearly all objects
from the statistical COMBO-17 sample (m$_r\le 23.6$); b) avoids
spurious deblending of the largest galaxies, which show ample
substructure in the HST images, reflecting spiral arms, OB
associations, \etc ; c) provides a homogeneous, flux- and
surface brightness-limited catalog of all sources in the \zb ~GEMS
mosaic, regardless of COMBO-17 or other external information.

\begin{figure}
\plotone{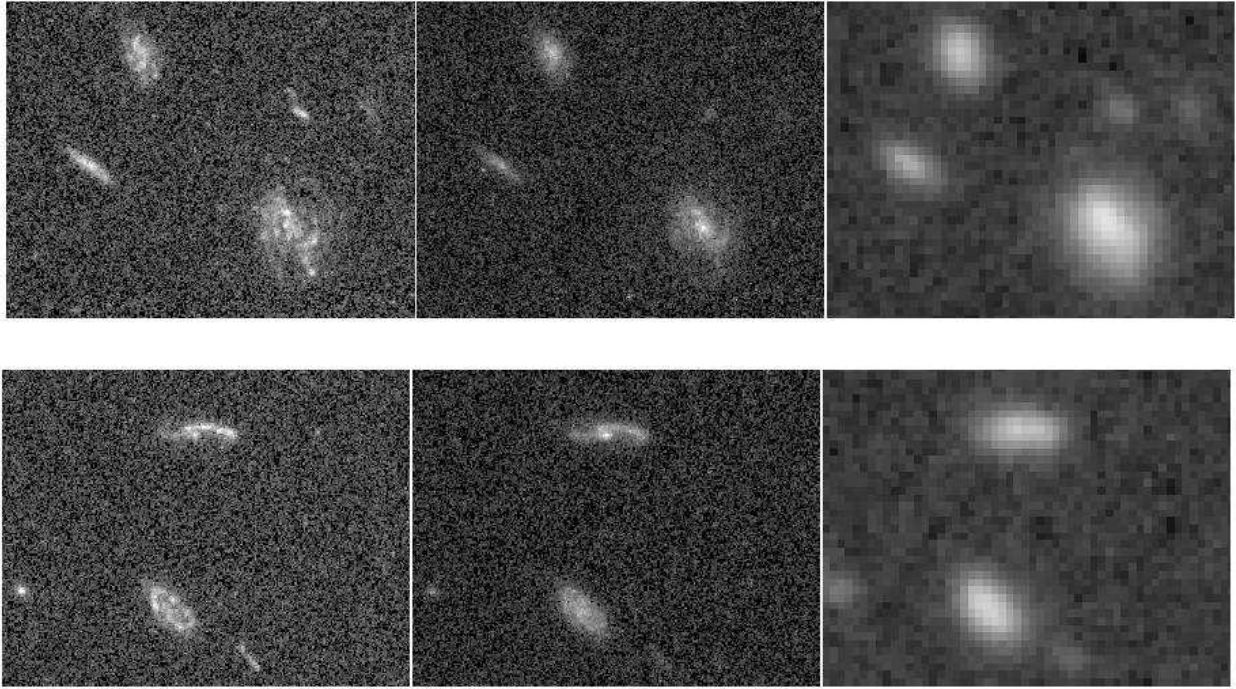} \caption{Comparison of the GEMS
V-band (F606W) data (left panel in each row), and z-band (F850LP) data (center)
with the deep, good seeing (0.7$^{\prime\prime}$ resolution) COMBO-17 
R-band data (right half of each sub-panel). While comparable point-source and
surface brightness sensitivity can be reached from the ground, the
advantage in source parsing and in the assessment of morphological
and structural information is manifest. Each panel is
$14^{\prime\prime}\times 10^{\prime\prime} $  on a side.
\label{GEMS-COMBO17}}
\end{figure}

\begin{figure}
\plotone{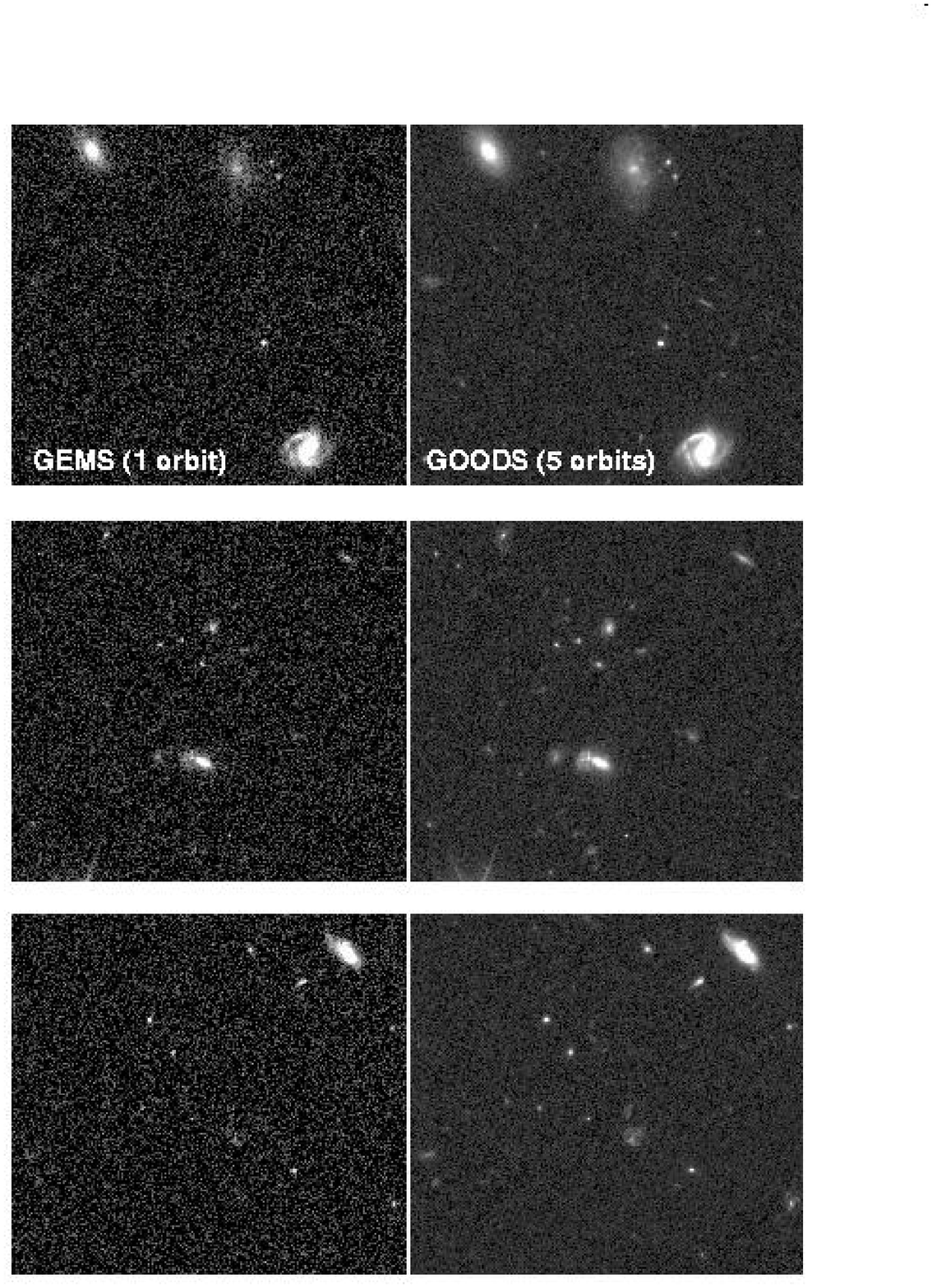} \caption{Comparison of a single-orbit
exposure in z-band, as used throughout GEMS, and of a 5-orbit
exposure in the same band, reflecting the full exposure time of the GOODS deep
imaging. Each panel is $33.6^{\prime\prime}x31^{\prime\prime}$ on a side, or 
0.00036 times the total GEMS mosaic area.\label{deep-shallow}}
\end{figure}

Even the first two requirements cannot be achieved with a single
SExtractor parameter setting. The point-source flux limit of the
GEMS \zb ~image is more than two magnitudes fainter than the
COMBO-17 catalog limit (for typical SEDs). But the ground-based
data, drawing on long exposures and large pixels, have at least as
high a surface brightness sensitivity as the ACS data (see
Fig.~\ref{GEMS-COMBO17}). To pick up all diffuse, low-surface
brightness objects from COMBO-17, the SExtractor program requires
an object detection threshold that is so sensitive that inevitably
weak features in the outer parts of large galaxies get deblended
spuriously, as illustrated in Figure 4. The judgment of
``over-deblending" was made by visual inspection, independently by
two of the co-authors (DHM \& MB).

To meet our catalog requirements, we then employ two different source
detection configurations with SExtractor: a) a conservative, {\it
cold} setting that avoids spurious deblending of large objects
with strong substructure, but does not pick up all faint,
low-surface-brightness objects in the COMBO-17 catalog; and b) a
{\it hot} version, assured to detect all faint objects at the
expense of an occasional over-deblending of a large source (see
Fig.~\ref{GEMS-COMBO17}). 
The following SExtractor configuration parameters define cold and hot
source detections:
(i) the detection threshold above background DETECT\_THRESH =
$2.3\sigma_{\rm bkg}$ (cold), $1.65\sigma_{\rm bkg}$ (hot);
(ii) the minimum number of connected pixels above threshold
DEBLEND\_MINAREA = 100 (cold), 45 (hot);
(iii) the minimum contrast between flux peaks for deblending multiple
sources DEBLEND\_MINCONT = 0.065 (cold), 0.06 (hot); and
(iv) the number of sub-thresholds considered during deblending
DEBLEND\_NTHRESH = 64 (cold), 32 (hot).
For both cases we employ a weight map ($\propto {\rm var}^{-1}$) and
a 3 pixel (FWHM) tophat filtering kernel.  The use of weight maps
reduces the number of spurious detections in low signal-to-noise
(S/N) areas of each image (e.g. near image edges).

 We apply SExtractor to our $z$-band
mosaic only; galaxies appear to have more substructure in the
bluer $V$-band imaging, which increases the number of spurious
over-deblendings and hampers meeting our detection criteria.  Our
final catalog consists of all (18,528) objects detected with our
cold configuration, augmented by an additional (23,153) hot
detections, but only those found {\it outside} of the isophotal
area of sources from the ``cold" catalog.  Note, that while the
SExtractor parameters for the hot and cold configurations were
fine tuned interactively, the final GEMS catalog (of 41,681 unique
sources) is produced strictly algorithmically.  This GEMS
catalog will be published and described in complete detail in C04.

We cross-correlate the final GEMS source catalog with the
ground-based COMBO-17 redshift catalog solely on the basis of the
object coordinates. There are 9,833 objects with
redshifts from COMBO-17 (m$_R\ltorder 24$) 
in the E-CDFS field of COMBO-17.  We consider as a
match the nearest redshift coordinate of a GEMS $z$-band position
if it is within $0.75\arcsec$.  The average RMS angular separation
between matches is $0.13\arcsec$ and the fraction of unclear or
blended detections at COMBO-17 coordinates is $\sim1 \%$.  We find
8,312 unique GEMS sources with redshift matches resulting in a 84,5\%
success rate; the roughly 14.5\% COMBO-17 objects without GEMS
detections are due to the larger coverage of the E-CDFS region by
COMBO-17 compared to our ACS imaging (Figure~\ref{gems-layout}).

\subsection{Image Simulations}

We explore the detection limits of the GEMS mosaic, with the above
described SExtractor cold$+$hot object detection configurations,
by extensive Monte-Carlo simulations (H\"au\ss ler \etal 2004,
{\it in prep}). Simulated galaxy images were added to the actual
data frames and processed as above. The detectability - and the
subsequent ability to extract structural parameters - depends
mostly on the effective surface brightness of the object
(Figures~5 and 6), with much weaker dependences on
the overall size and the axis ratio (H\"au\ss ler \etal 2004,
{\it in prep}).
When defining the mean surface brightness of a galaxy image as:
$$\langle \mu_z \rangle\equiv
\frac{ I_{tot}}{2}\times \frac{1}{\pi r_{eff}^2\times q},$$ where
$I_{tot}$ is the total flux, $r_{eff}$ is the effective,
or half-light  radius and $q$ is the
axis ratio, 
we find the characteristic (80\%) completeness limit of the GEMS {\it
galaxy} sample to be $\langle \mu_z\rangle=24$ for exponential
profiles and $\langle \mu_z\rangle=25$ mag arcsec$^{-2}$
 for R$^{1/4}$ profiles.

\subsection{Point Spread Function}

While the majority of galaxies in the combined COMBO-17/GEMS
sample are resolved in the sense that the intrinsic half-light
radius, R$_e$, is larger than the diffraction limit of HST
($\lambda / D\approx  0.077\asec$ at \zb ), virtually all objects
have central flux gradients on angular scales much smaller than
the point spread function's (PSF) FWHM.
 This is particularly true in the cases of AGNs,
where the unresolved central source often dominates. Any image
modeling of the galaxies in GEMS requires therefore an accurate
knowledge of the PSF.

There are two basic ways to construct a model PSF for the
subsequent image interpretation: a calculation of the
theoretically predicted PSF, using \eg TinyTim (Krist 1993) or an
empirical construction from the point sources within the data (\eg
{\it daophot}, Stetson \etal 1987). By necessity, the synthetic
PSF is based on model assumptions, some of which are not
sufficiently understood. On the other hand, the empirical PSF
depends on a finite number of bright but unsaturated stars in the
data. For strong PSF variations within the field, they may be hard
to derive from fields at high Galactic latitude.

The large number of ACS images obtained within GEMS, homogeneously
acquired and reduced, has enabled us to study the inter- and
intra-tile variations of the PSF in some detail. We found that variations among different
tiles are negligible, while the PSF dependence on position within
an ACS frame is noticeable but still small. We have performed
extensive simulations assuring us that for fitting galaxy images
without prominent AGN components, one universal, empirical high
S/N PSF per filter is fully sufficient for all tiles and for all
positions within each tile (H\"au\ss ler \etal 2004, {\it in prep.}).

Active Galaxies with a strong nuclear point source require a more
elaborate treatment, due to the spatial PSF variations within the 
tiles. For such cases we use appropriate sub-tile PSF
representations, jointly derived from all pointings. For each AGN 
the sub-tile PSF is constructed from at least $\sim 30$ stellar
images near its pixel position (Jahnke \etal 2004, {\it in
prep.}).

\section{Science Analysis Plan}

The science analysis plan is obviously evolving, as we learn about the
potential and limitations of the data, and as we learn from the data themselves.Here, we provide an outline of the current efforts:

\bigskip\noindent{\sl Galaxy Fitting}

Quantifying the radial distribution, shape, and
morphology of galaxy images with a small number of parameters is a
challenging problem with a long history.
No single simplified image description can fit all science goals
and we have therefore chosen a multi-pronged approach.

First, one can try to specify model-independent
numbers, such as the magnitude within a set of fiducial apertures,
or the effective (= half light) radius, or the luminosity-weighted
mean ellipticity of the image. This requires a PSF deconvolution
and proper noise weighting. Second, one can fit a parameterized,
usually two-dimensional galaxy model to the image portions. Such
models can be single component, \eg exponential disks (Freeman 1970),
de Vaucouleurs (1959) models, or the more general S{\' e}rsic (1968)
models, or they can be multi-component models, \eg representing a
bulge, a disk, and perhaps an active nucleus.
We fit the galaxy images with two different codes for
parameterized models: GALFIT (Peng \etal 2002) and GIM2D (Simard
\etal 1999).

On the other hand, non-axisymmetric components such as stellar bars
and spiral arms, as well as asymmetries, tidal tails, and other
distorted features do not lend themselves  to a parameterized model
description. In order to identify stellar bars and characterize  their
properties  such as ellipticities and sizes, we perform isophotal fits
to galaxy images (Jogee \etal 2004 {\it in prep.}). These isophotes
provide a  guide to the underlying orbital structure of the bar and disk,
with the bar leading to a characteristic peak in ellipticity  over
a plateau in position angle   (e.g,  Wozniak et al. 1995; Friedli
et al 1996; Jogee \etal  2002) over the region dominated by
the $x_1$  family of stellar orbits (Contopoulos \& Papayannopoulos 1980).
The resulting profiles of surface brightness and ellipticity  are
deprojected to derive intrinsic bar properties.

Furthermore,  we use the CAS code  (Conselice, Bershady, \& Jangren 2000)
to quantify the asymmetry  {\it A}  and   concentration  {\it C}
of the images. We measure {\it A} in the rest-frame $B$-band in
order to trace  distorted features such as prominent tails, arcs,
and double nuclei characteristic of interacting and merging systems,
and thereby constrain  the merger/interaction history (Jogee \etal
2004 {\it in prep.}. These techniques are augmented by a qualitative,
by-eye classification for a subset (Bell \etal 2003b).
The resulting structural parameters will then be used to study
the luminosity--size, mass-size, compactness-color (Bell \etal 2003b),
relations and their redshift evolution.

A detailed comparison and test of these approaches is underway
(H\"au\ss ler \etal 2004, {\it in prep}).

A separate fitting effort is underway to explore the host galaxies for all
known AGNs in the GEMS field (\eg Wolf \etal 2003b), requiring foremost
the removal of the nuclear component of the image. Compared to existing
host galaxy studies, the relatively faint nuclei, promise a very high
success rate in host galaxy detection.

\bigskip\noindent{GEMS and COMBO-17}

Beyond drawing on  COMBO-17 redshifts to
derive intrinsic luminosities and sizes, there are many
intriguing ways to combine the two data sets. One can define
``early-type" galaxies samples completely independently on the
basis of SED or on radial luminosity profile (high Sersic index)
and compactness, and compare to which extent the definitions lead
to the same sample as color selection (Bell \etal 2003c). One can
explore the evolution of the luminosity--size and
luminosity--(stellar) mass relations for the overall galaxy
population and for disks and spheroids separately. One can study
the evolution of the galaxy size function, i.e. the evolution of
the space density of galaxies with a given size. The
stellar masses estimated from SED modeling (Borch \etal 2003,
{\it in prep.}) of COMBO-17 data can be used to explore the structural
and star-formation properties of galaxies as a function of stellar
mass and of stellar surface-mass density.

\bigskip\noindent{Nearby Galaxy Comparison Sample}

Any study of galaxy evolution should be anchored at the present epoch.
At low redshifts ($z \ltorder 0.3$), the volume of GEMS is
``small", i.e. smaller than the galaxy correlation length. To
increase the effective redshift range over which to study evolution, it is
tantamount to assemble and analyze consistently a sample of
present-epoch galaxies.
Even though the data for the present-day universe have drastically improved,
e.g. through SDSS, local universe data need considerable tailoring  
to mesh optimally with GEMS. We have chosen the SDSS galaxy sample from
the DR1 (Abazijan \etal 2003) for this comparison.  Remarkably, only to distances of
$z\sim 0.03$ is the spatial
resolution of the SDSS imaging (at $\sim 1.2\asec$ angular
resolution) sufficient to match the ACS's resolution for objects
at $z\sim 0.75$. For this local comparison sample we are deriving 
SED-based
estimates of the stellar mass and U-V rest-frame color estimates for
consistency with the COMBO-17 data on the GEMS field.

\bigskip\noindent{Model Comparison and Follow-Up}

Initial comparison between GEMS data and cosmological models will
focus on semi-analytic models (Somerville \etal {\it in prep.})
that incorporate predictions for the sizes of disks and spheroids.
Owing to the computational expense of N-body/SPH simulations,
semi-analytic models are the only ones that can predict population
statistics of structural parameters. 

Modeling of the colors in terms of stellar populations, and
estimating \eg mass-to-light ratios will build on COMBO-17's multi-band SED
modeling (Borch \etal 2003). With GEMS such an analysis can be
done pixel-by-pixel, using observations in two bands and 
following the approach of  Bell \& de~Jong (2001). This analysis will
ultimatively allow to estimate e.g. half-mass, rather than half-light,
radii for the stellar bodies of galaxies.

To understand better the stellar populations of the galaxies in
the GEMS field, to increase the redshift range over which galaxies
can be studied, to improve the stellar mass estimates and star
formation rate estimates, a suite of analyses is underway within
the GEMS collaboration and by others. This includes near-IR
imaging (\eg Chen \etal 2003) and X-ray imaging with Chandra
(Cycle 5, PI: N. Brandt). With this follow-up, the data from GEMS seem
certain to be scientifically fruitful beyond the project's
initially articulated goals.

\section{Summary}

The GEMS project, which stands for ``Galaxy Evolution from Morphologies and SEDs'',
has produced the largest color image taken with HST to date, providing 
structural and morphological information for over 10,000 distant galaxies.
As an overarching summary and reference for a series of papers with individual 
results from GEMS, we have presented here an overview of the science goals, 
the experiment
layout, the observations, the main data reduction steps and the initial 
data modeling
philosophy.

While the core science goals, mapping the evolution of galaxy bulges, 
disks and mergers
over the last half of the universe's age, can draw on a single data 
processing approach, described
here, the investigation of AGN host galaxies, of gravitational lensing, 
and of internal color 
gradients in galaxies will require further processing.
To ensure broad scientific impact of the GEMS data, we will provide all 
processed data, 
along with redshifts for 10,000 galaxies, via the World Wide Web in
Spring 2004.

\acknowledgments

We are grateful to Alison Vick and Guido de~Marchi for 
support in preparing the HST observations. This research
was supported by STScI through HST-GO-9500.01. 
Support for the GEMS project was provided by NASA through grant number GO-9500
from the Space Telescope Science Institute, which is operated by the
Association of Universities for Research in Astronomy, Inc. for NASA, under
contract NAS5-26555.
EFB and SFS ackowledge financial support provided through
the European Community's Human Potential Program under contract
HPRN-CT-2002-00316, SISCO (EFB) and HPRN-CT-2002-00305, Euro3D RTN (SFS).
CW was supported by the PPARC rolling grant in Observational Cosmology at
University of Oxford.
SJ acknowledges support from  the National Aeronautics
and Space Administration (NASA) under LTSA Grant  NAG5-13063
issued through the Office of Space Science.
D.\ H.\ M.\ acknowledges support from the National Aeronautics
and Space Administration (NASA) under LTSA Grant NAG5-13102
issued through the Office of Space Science.
HWR acknowledges support from the German-Israeli Science Foundation.

\clearpage

\clearpage

\clearpage

\clearpage

\begin{figure}
%\plotfiddle{fig4.eps}{7.2truein}{0}{1.0}{1.0}{80}{0}
\vbox to7.6in{\rule{0pt}{7.2in}}
\includegraphics{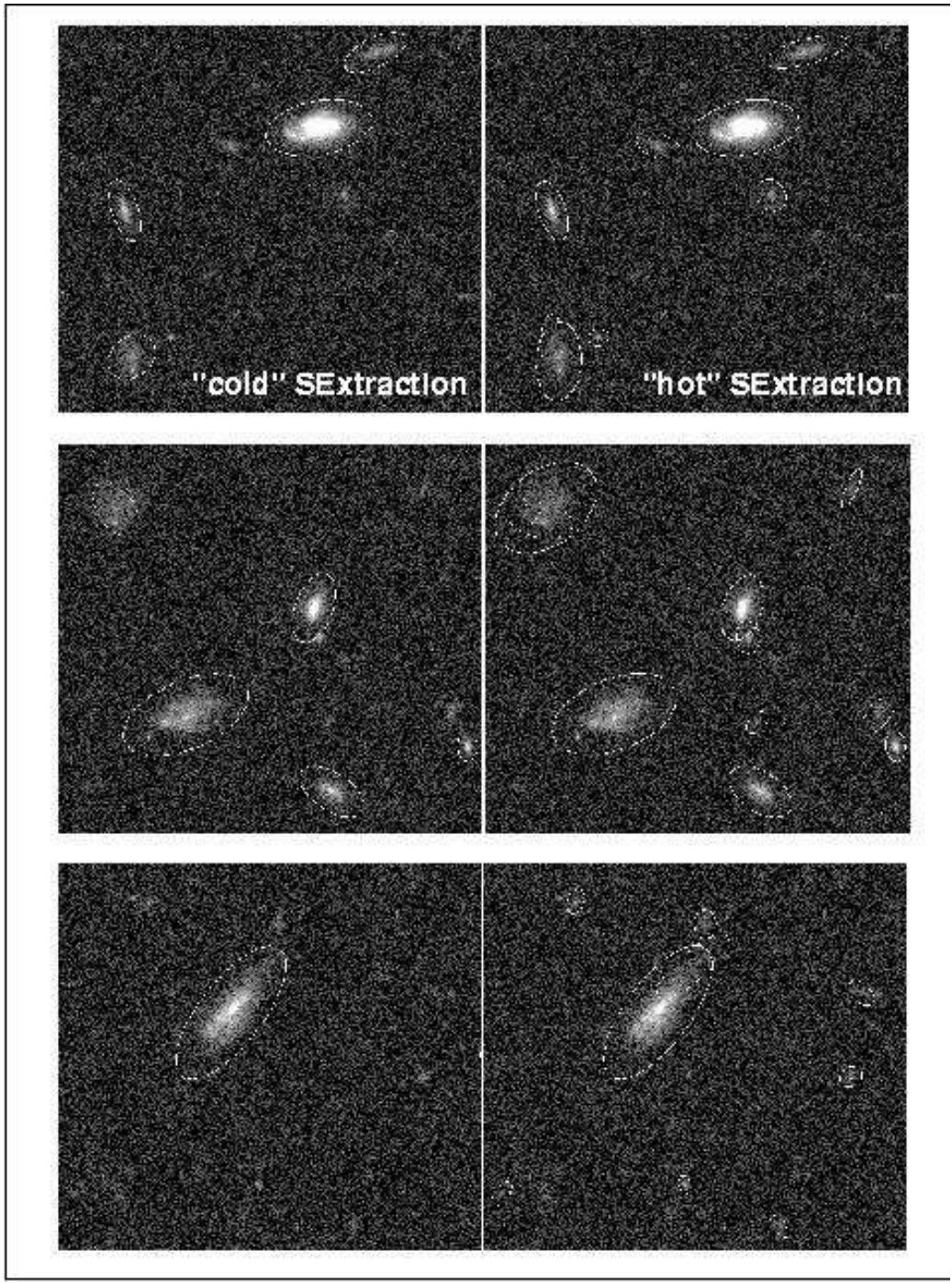}
\caption{Two-pass strategy for object detection and deblending:
the left-subpanels show the source identification with a
conservative (``cold") setting of the SExtractor parameters that
avoid over-deblending of large objects and galaxies with
lumpy structure. The right panels shows the result of SExtractor
with a parameter setting that picks up objects closer to the noise
threshold, at the expense of occasionally breaking up objects
erroneously. Our final object catalog consists of the left, cold
objects augmented by the missing right, ``hot" objects, but only 
those that do not overlap with a cold object. Each panel is $16.8^{\prime\prime}
\times 15.4^{\prime\prime}$ on a side.
\label{hot-cold-SExtraction}}
\end{figure}
\clearpage

\begin{figure}
\vbox to2.6in{\rule{0pt}{2.6in}}
\includegraphics{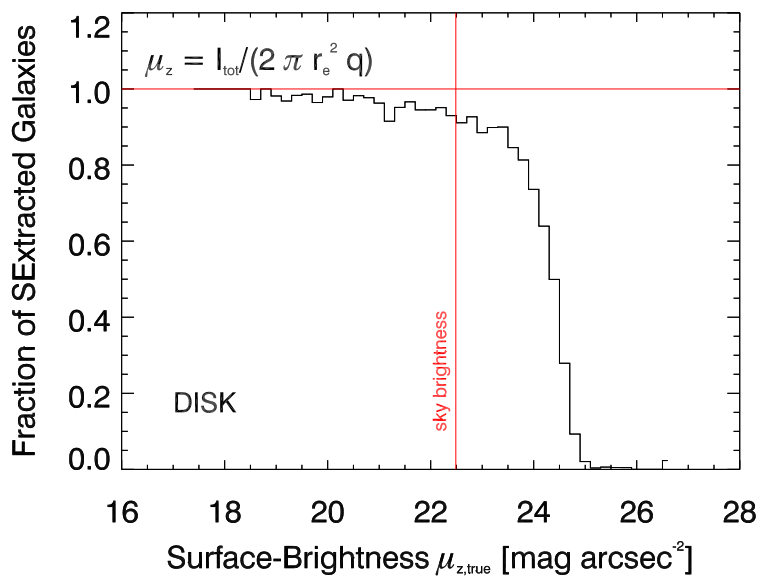} \caption{Completeness in surface brightness 
of the GEMS z-band imaging for
detecting exponential disks with SExtractor and subsequently 
fitting them.  The definition of the mean surface
brightness is given in \S 4.2. Even for bright galaxies, object
overlap causes a small fraction of them not to pass detection and fitting.
The vertical line indicates the background flux of the ACS data.}
\end{figure}

\clearpage

\begin{figure}
\vbox to2.6in{\rule{0pt}{2.6in}}
\includegraphics{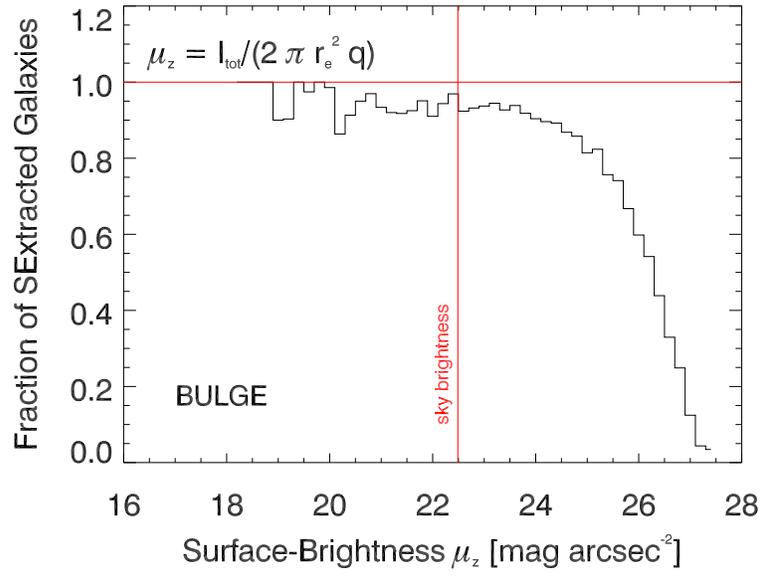}\caption{Completeness in surface brightness 
of the GEMS z-band imaging as in Fig 5., but for detecting and subsequently fitting
de~Vaucouleurs bulges. The higher central concentration permits the
detection of r$^{1/4}$ objects with lower effective surface
brightness than exponential disks. \label{bulge-detection}}
\end{figure}


\begin{thebibliography}{}


\bibitem[Abazajian \etal (2003)]{2003AJ....126.2081A} Abazajian, K., \etal
2003, \aj, 126, 2081 


\bibitem[Abraham \etal (1996)]{1996MNRAS.279L..47A} Abraham, R.~G., Tanvir,
N.~R., Santiago, B.~X., Ellis, R.~S., Glazebrook, K., \& van den
Bergh, S.\ 1996, \mnras, 279, L47


\bibitem[Bell \& de Jong(2001)]{2001ApJ...550..212B} Bell, E.~F., \& de
Jong, R.~S.\ 2001, \apj, 550, 212


\bibitem[Bell, McIntosh, Katz, \& Weinberg(2003a)]{2003ApJ...585L.117B} 
Bell, E.~F., McIntosh, D.~H., Katz, N., \& Weinberg, M.~D.\ 2003a, 
ApJS, {\it in press} (astro-ph/0302543).


\bibitem[Bell \etal (2003b)]{bell03b}
Bell, E.~F., Wolf, C., Meisenheimer, K., Rix, H.-W., Borch, A., Dye, S.,
Kleinheinrich, M., \& McIntosh, D.~H.\ 2003b astro-ph/0303394

\bibitem[Bell \etal (2003c)]{bell03c}
Bell, E.~F., et al. 2003c, submitted to ApJL (astro-ph/0308272)

\bibitem[Bertin \& Arnouts(1996)]{bertin96} Bertin, E., \& Arnouts, S.\ 
1996, \aaps, 117, 39


\bibitem[Blanton \etal (2003c)]{blanton03}
        Blanton, M.~R., \etal  2003c, submitted to \apj
        { }(astro-ph/0210215)


\bibitem[Borch, Meisenheimer, Wolf, \& Gray(2003)]{2003Ap&SS.284..965B}
Borch, A., Meisenheimer, K., Wolf, C., \& Gray, M.\ 2003, \apss,
284, 965


\bibitem[Chen \etal (2003)]{2003ApJ...586..745C} Chen, H., \etal 2003,
\apj, 586, 745



\bibitem[Coil \etal (2003)]{DEEP-clustering} Coil, A., \etal 2003, ApJ, 
{\it submitted} (astro-ph/0305586)


\bibitem[Cole, Lacey, Baugh, \& Frenk(2000)]{cole-semi-analytic} Cole, S.,
Lacey, C.~G., Baugh, C.~M., \& Frenk, C.~S.\ 2000, \mnras, 319,
168


\bibitem[Colless \etal (2001)]{colless01}
        Colless, M., \etal  2001, \mnras, 328, 1039


\bibitem[Conselice, Bershady, \& Jangren(2000)]{2000ApJ...529..886C}
Conselice, C.~J., Bershady, M.~A., \& Jangren, A.\ 2000, \apj,
529, 886

\bibitem[Contopoulos, G. \& Papayannopoulos, T. (1980)] {}
Contopoulos, G. \& Papayannopoulos, T. 1980, A\&A, 92, 33

\bibitem[Daddi et al.(2001)]{2001A&A...376..825D} Daddi, E., Broadhurst, 
T., Zamorani, G., Cimatti, A., R{\" o}ttgering, H., \& Renzini, A.\ 2001, 
\aap, 376, 825 


\bibitem[Davis et al.(2003)]{2003SPIE.4834..161D} Davis, M.~et al.\ 2003, 
\procspie, 4834, 161 


\bibitem[de Jong(1996)]{1996A&A...313...45D} de Jong, R.~S.\ 1996, \aap,
313, 45


\bibitem[de Vaucouleurs(1959)]{DeVauc} de~Vaucouleurs, G.\ 1959, 
{\it Handbuch der Physik}, 53, 311.


\bibitem[Dickinson, Giavalisco, \& The Goods
Team(2003)]{2003mglh.conf..324D} Dickinson, M., Giavalisco, M., \&
The Goods Team 2003, The Mass of Galaxies at Low and High
Redshift.~Proceedings of the ESO Workshop held in Venice, Italy,
24-26 October 2001, 324


\bibitem[Ellis \etal (1996)]{autofib} Ellis, R.~S., Colless,
M., Broadhurst, T., Heyl, J., \& Glazebrook, K.\ 1996, \mnras,
280, 235


\bibitem[Ferguson \etal (2003)]{Ferguson} Ferguson, H. \etal 2003,
astro-ph/0309058.

\bibitem[Ford \etal (2003)]{2003SPIE.4854...81F} Ford, H.~C., \etal 2003,
\procspie, 4854, 81

\bibitem[Freeman(1970)]{1970ApJ...160..811F} Freeman, K.~C.\ 1970, \apj, 
160, 811 

\bibitem[Fried \etal (2001)]{2001A&A...367..788F} Fried, J.~W., \etal
2001, \aap, 367, 788


\bibitem[Friedli \etal (1996)]{}
Friedli, D., Wozniak, H., Rieke, M., \& Bratschi, P. 1996,
A\&AS, 118, 461

\bibitem[Giacconi \etal (2001)]{2001ApJ...551..624G} Giacconi, R., \etal
2001, \apj, 551, 624

\bibitem[Giavalisco et al.(1998)]{1998ApJ...503..543G} Giavalisco, M., 
Steidel, C.~C., Adelberger, K.~L., Dickinson, M.~E., Pettini, M., \& 
Kellogg, M.\ 1998, \apj, 503, 543 

\bibitem[Giavalisco \& GOODS Team(2003)]{2003AAS...202.1703G} Giavalisco,
M.~\& GOODS Team 2003, American Astronomical Society Meeting, 202,


\bibitem[Im \etal (2002)]{2002ApJ...571..136I} Im, M., \etal 2002, \apj,
571, 136

\bibitem[Jogee \etal 2002]{}
Jogee, S., Knapen, J. H.,  Laine, S., Shlosman, I., Scoville,
N. Z., \&  Englmaier, P. 2002, ApJL, 570, L55  (astro-ph/0201208)

\bibitem[Katz \& Gunn(1991)]{1991ApJ...377..365K} Katz, N.~\& Gunn, J.~E.\ 
1991, \apj, 377, 365


\bibitem[Kauffmann, White, \& Guiderdoni(1993)]{1993MNRAS.264..201K}
Kauffmann, G., White, S.~D.~M., \& Guiderdoni, B.\ 1993, \mnras,
264, 201


\bibitem[Kauffmann \etal (2003)]{2003MNRAS.341...33K}
Kauffmann, G., \etal 2003, \mnras, 341, 33

\bibitem[Kranz, Slyz, \& Rix(2003)]{2003ApJ...586..143K} Kranz, T., Slyz,
A., \& Rix, H.-W.\ 2003, \apj, 586, 143


\bibitem[Krist(1993)]{1993adass...2..536K} Krist, J.\ 1993, ASP Conf.~Ser.~
52: Astronomical Data Analysis Software and Systems II, 2, 536


\bibitem[Labb{\' e} \etal (2003)]{2003ApJ...591L..95L} Labb{\' e}, I., \etal 
2003, \apjl, 591, L95 


\bibitem[Le F{\` e}vre \etal (2003)]{2003Msngr.111...18L} Le F{\` e}vre,
O., \etal 2003, The Messenger, 111, 18


\bibitem[Lenzen \etal (2003)]{2003SPIE.4841..944L} Lenzen, R., \etal 2003,
\procspie, 4841, 944


\bibitem[Lilly, Le F{\` e}vre, Hammer, \& Crampton(1996)]{1996ApJ...460L...1L}
Lilly, S.~J., Le F{\` e}vre, O., Hammer, F., \& Crampton, D.\ 
1996, \apjl, 460, L1


\bibitem[Lilly \etal (1998)]{1998ApJ...500...75L} Lilly, S., \etal 1998,
\apj, 500, 75

\bibitem[Moustakas\etal (2003)]{Moustakas} Moustakas L. \etal 2003,
astro-ph/0309187.


\bibitem[Mutchler \etal (2003)]{multidrizzle} Mutchler, M., Koekemoer, A, 
\& Hack, W.\ 2003,
2002 HST Calibration Workshop, STScI, 2002, {\it eds.} Arribas, S.,
\etal

\bibitem[Norberg \etal (2002)]{2002MNRAS.336..907N} Norberg, P., \etal 
2002, \mnras, 336, 907 

\bibitem[Peng, Ho, Impey, \& Rix(2002)]{2002AJ....124..266P} Peng, C.~Y.,
Ho, L.~C., Impey, C.~D., \& Rix, H.-W.\ 2002, \aj, 124, 266


\bibitem[Percival \etal (2002)]{2002MNRAS.337.1068P} Percival, W.~J., \etal 
2002, \mnras, 337, 1068


\bibitem[Phleps \& Meisenheimer(2003)]{2003A&A...407..855P} Phleps, S.~\& 
Meisenheimer, K.\ 2003, \aap, 407, 855 


\bibitem[Rix \& Rieke (1993)]{1993ApJ...418..123R} Rix, H.-W., \& Rieke, 
M.~J.\ 1993,
\apj, 418, 123


\bibitem[Rosati \etal (2002)]{2002ApJ...566..667R} Rosati, P., \etal 2002,
\apj, 566, 667


\bibitem[Sersic(1968)]{1968adga.book.....S} S{\' e}rsic, J.~L.\ 1968, Cordoba, 
Argentina: Observatorio Astronomico, 1968


\bibitem[Shen \etal (2003)] {2003MNRAS.343.978} Shen, S., \etal 2003, 
\mnras, 343, 978


\bibitem[Simard \etal (1999)]{1999ApJ...519..563S} Simard, L., \etal 1999,
\apj, 519, 563


\bibitem[Simard et al.(2002)]{2002ApJS..142....1S} Simard, L.~et al.\ 2002, 
\apjs, 142, 1 


\bibitem[Skrutskie \etal (1997)]{skrut}
        Skrutskie, M.~F., \etal  1997, in `The Impact of
        Large Scale Near-IR Sky Surveys', eds.~F.~Garzon, \etal
	(Dordrecht: Kluwer Academic Publishing Company), 25 


\bibitem[Somerville \& Primack(1999)]{somerville-semi-analytic} Somerville,
    R.~S.~\& Primack, J.~R.\ 1999, \mnras, 310, 1087


\bibitem[Spergel \etal (2003)]{WMAP} Spergel, D., \etal 2003, ApJ, 
{\it in press}


\bibitem[Springel, Yoshida, \& White(2001)]{2001NewA....6...79S} Springel,
V., Yoshida, N., \& White, S.~D.~M.\ 2001, New Astronomy, 6, 79


\bibitem[Steidel \etal (1996)]{ly-break} Steidel, C.~C.,
Giavalisco, M., Pettini, M., Dickinson, M., \& Adelberger, K.~L.\ 
1996, \apjl, 462, L17


\bibitem[Steinmetz \& Navarro(2002)]{2002NewA....7..155S} Steinmetz, M.~\&
Navarro, J.~F.\ 2002, New Astronomy, 7, 155


\bibitem[Stetson(1987)]{1987PASP...99..191S} Stetson, P.~B.\ 1987, \pasp,
99, 191


\bibitem[Strateva \etal (2001)]{2001AJ....122.1861S} Strateva, I., \etal 
2001, \aj, 122, 1861 


\bibitem[van Dokkum, Franx, Kelson, \&
Illingworth(1998)]{1998ApJ...504L..17V} van Dokkum, P.~G., Franx,
M., Kelson, D.~D., \& Illingworth, G.~D.\ 1998, \apjl, 504, L17


\bibitem[Williams \etal (1996)]{hdfn} Williams, R.~E., \etal 1996, \aj, 
112, 1335


\bibitem[Wolf \etal (2001)]{2001A&A...377..442W} Wolf, C., Dye, S.,
Kleinheinrich, M., Meisenheimer, K., Rix, H.-W., \& Wisotzki, L.\ 
2001, \aap, 377, 442


\bibitem[Wolf \etal (2003a)]{2003A&A...401...73W} Wolf, C., Meisenheimer, K.,
Rix, H.-W., Borch, A., Dye, S., \& Kleinheinrich, M.\ 2003a, \aap,
401, 73


\bibitem[Wolf \etal (2003b)]{astro/ph-0304072} Wolf, C., Wisotzki, L., 
Borch, A., Dye, S.,
Kleinheinrich, M., Meisenheimer, K.\ 2003b, A\& A, {\it in press}
astro-ph/0304072

\bibitem[Wozniak \etal 1995]{}
Wozniak, H., Friedli, D., Martinet, L., Martin, P., Bratschi, P.
1995, A\&AS 111, 115.



\bibitem[York \etal (2000)]{york00}
        York, D.~G., \etal 2000, \aj, 120, 1579
\end{thebibliography}
\end{document}